\documentclass[12pt,fleqn]{article}

\usepackage{amsmath,amssymb,amsfonts,graphicx,epsfig}
\def\bea{\begin{eqnarray}}
\def\eea{\end{eqnarray}}
\def\be{\begin{equation}}
\def\ee{\end{equation}}
\def\ba{\begin{array}}
\def\ea{\end{array}}
\def\nn{\nonumber}

\def \gsim{\mathrel{\vcenter
{\hbox{$>$}\nointerlineskip\hbox{$\sim$}}}}

\DeclareFontFamily{OT1}{pzc}{}
\DeclareFontShape{OT1}{pzc}{m}{it}{<-> s * [1.4] pzcmi7t}{}
\DeclareMathAlphabet{\mathpzc}{OT1}{pzc}{m}{it}
\DeclareMathAlphabet{\mathscr}{OT1}{pzc}{m}{it}

\def\a{&\hspace{-7pt}}
\def\b{&\hspace{-10pt}}

\begin{document}
\pagestyle{plain}

\renewcommand{\theequation}{\arabic{section}.\arabic{equation}}
\setcounter{page}{1}

\begin{titlepage} 
\begin{center}

$\;$

\vskip 0.5 cm

\begin{center}
{\LARGE{ \bf Metastable spontaneous breaking \\[2mm] of N=2  supersymmetry}}
\end{center}

\vskip 1cm

{\large 
Beno\^it L\'egeret, Claudio A. Scrucca and Paul Smyth
}

\vskip 0.5cm

{\it
Institut de Th\'eorie des Ph\'enom\`enes Physiques, EPFL, \\
\mbox{CH-1015 Lausanne, Switzerland}\\
}

\vskip 1.5cm

\begin{abstract}

We show that contrary to the common lore it is possible to spontaneously break 
N=2 supersymmetry even in simple theories without constant Fayet-Iliopoulos terms. 
We consider the most general N=2 supersymmetric theory with one hypermultiplet 
and one vector multiplet without Fayet-Iliopoulos terms, and show that metastable 
supersymmetry breaking vacua can arise if both the hyper-K\"ahler and the 
special-K\"ahler geometries are suitably curved. We then also prove 
that while all the scalars can be massive, the lightest one is always lighter than the 
vector boson. Finally, we argue that these results also directly imply that metastable 
de Sitter vacua can exist in N=2 supergravity theories with Abelian gaugings and no 
Fayet-Iliopoulos terms, again contrary to common lore, at least if the cosmological 
constant is sufficiently large.

\end{abstract}

\bigskip

\end{center}
\end{titlepage}

\newpage

\section{Introduction}

It is by now well understood that the difficulty of achieving metastability 
for vacua leading to spontaneous supersymmetry breaking has a simple and 
universal origin related to Goldstone's theorem applied to supersymmetry. 
Indeed, since the Goldstino fermion must be massless, its sGoldstini scalar
superpartners have masses that are entirely controlled by supersymmetry breaking 
effects and cannot be adjusted through supersymmetric mass terms. More precisely,
it turns out that the average mass of these sGoldstini is entirely controlled by the 
geometry of the scalar manifold and the data of the local gauge symmetries, if present
\cite{GRS1,GRS2,GRS3} (see also \cite{DD}).
This fact is true not only in rigid supersymmetry but also in local supersymmetry,
and its consequences have already been extensively investigated in a number of 
situations \cite{CGRGLPS1,CGRGLPS2}. The general outcome is that one can infer
a simple upper bound on the mass of the lightest scalar, which depends only on 
the geometric data of the theory and is increasingly stringent in theories with 
increasing number of supercharges. In some special classes of theories, this upper 
bound is vanishing or even negative, and therefore results in no-go theorems 
forbidding metastable spontaneous supersymmetry breaking \cite{GRLS, MANY}.

In N=1 theories, the situation is quite simple and clear, especially in the rigid limit \cite{BS}. 
In theories with only chiral multiplets, the upper bound on the lightest mass is controlled 
by the sectional curvature of the scalar manifold along the supersymmetry breaking direction. 
This implies that in renormalizable theories with flat geometry, there are always two massless 
scalars (corresponding to the so-called pseudo-moduli of O'Raifeartaigh models), while in 
non-renormalizable non-linear sigma-models all the scalars can be massive if the sectional 
curvature can be positive. In theories involving also vector multiplets, the situation is qualitatively 
similar but the vector multiplets can give quantitatively important effects. As a result, all the scalars 
can in general be massive, already in renormalizable theories (corresponding to the absence of 
pseudo-moduli in gauged O'Raifeartaigh models) and clearly also in non-renormalizable ones,
even without Fayet-Iliopoulos terms.

In N=2 theories, the situation is more interesting and less clear, even in the rigid limit \cite{JS}. 
In theories with only hypermultiplets, supersymmetry breaking stationary points are possible 
only for curved geometries and the upper bound on the lightest mass happens to always vanish, 
as a consequence of the structure of the sectional curvature of hyper-K\"ahler manifolds. This 
implies that there is always at least one tachyonic scalar, and thus that the vacuum cannot be 
metastable. In theories with only Abelian vector multiplets, a similar result holds true, and again 
it is impossible to get metastable supersymmetry breaking vacua. On the other hand, in more 
general theories with both hyper- and vector multiplets, only partial results concerning the possible 
metastability of supersymmetry breaking vacua exist. One general result in this direction has been 
presented in \cite{AB}, where it was shown that for theories admitting an $SU(2)_R$ symmetry 
and a supercurrent conservation law involving a linear superconformal anomaly multiplet, 
it is impossible to construct a consistent non-linear realization of N=2 supersymmetry.
This suggests that in this class of theories there should be an unavoidable obstruction against 
having a non-supersymmetric stationary point at all or more plausibly against it to be metastable,
originating from the presence of the $SU(2)_R$ symmetry or the existence of the linear 
superconformal anomaly multiplet. The common lore is that in order to spontaneously break 
supersymmetry, one needs constant Fayet-Iliopoulos terms, which spoil the $SU(2)_R$ 
symmetry in the vector multiplet sector (see \cite{F,ADJ} for examples). However, the $SU(2)_R$
symmetry can also be spoiled by the lack of isometries on the scalar manifold in the hypermultiplet 
sector, and this might provide an alternative to Fayet-Iliopoulos terms.

The aim of this letter is to assess whether the possibility of having metastable supersymmetry 
breaking in N=2 theories is really linked to the presence of Fayet-Iliopoulos terms. 
For this we shall study in full generality the simplest class 
of such theories for which no-go theorems based on the sGoldstino masses do not exist so far, 
namely theories involving just one hypermultiplet and one vector multiplet with an Abelian
gauge symmetry.

\section{General setup}

Let us consider the most general N=2 supersymmetric theory involving only one hypermultiplet and 
one vector multiplet without Fayet-Iliopoulos terms. This has the form of a gauged 
non-linear sigma-model on a target space that is the product of an arbitrary four-dimensional
hyper-K\"ahler manifold admitting a triholomorphic isometry and an arbitrary two-dimensional 
special-K\"ahler manifold. The bosonic part of the action, describing the four real 
scalars $q^u$ belonging to the hypermultiplet and the complex scalar $z$ plus the 
real vector $A_\mu$ belonging to the vector multiplet, is given by \cite{AGF1,AGF2,HKLR,
dWvP,dWLvP,S,DFF,BW} (see \cite{JS} for a recent review on the rigid case):
\bea
{\cal L} = - \text{\small $\frac 14$} \rho \hspace{1pt} F_{\mu \nu} F^{\mu \nu} 
+ \text{\small $\frac 14$} \theta F_{\mu \nu} \tilde F^{\mu \nu} 
- \text{\small $\frac 12$} \hspace{1pt} g_{uv}\hspace{1pt} D_\mu q^u D^\mu q^v 
- g_{z \bar z}\, \partial_\mu z\, \partial^\mu \bar z
- V
\eea
In this expression $F_{\mu \nu} = \partial_\mu A_\nu - \partial_\nu A_\mu$, 
$\tilde F_{\mu \nu} = \frac 12\, \epsilon_{\mu \nu \rho \sigma} F^{\rho \sigma}$ and
$D_\mu q^u = \partial_\mu q^u + k^u A_\mu$. Moreover, $g_{uv}$ denotes 
the metric of the hyper-K\"ahler manifold, $k^u$ a triholomorphic Killing vector on it
and $P^i$ the three associated Killing potentials. Similarly, $g_{z \bar z}$ denotes the 
metric of the special-K\"ahler manifold, while $\rho$ and $\theta$ are the real 
and imaginary parts of the corresponding holomorphic gauge kinetic function, 
such that in particular $\rho = g_{z \bar z}$. Finally
\bea
V = g_{uv} k^u k^v |z|^2 + \text{\small $\frac 12$} \hspace{1pt} \rho^{\text{--}1} |\vec P|^2 \,.
\eea
Happily, it turns out that there exists a general local parametrization for the two kinds 
of manifolds that are involved in this construction, in terms of two harmonic functions 
$f$ and $l$ of three and two real variables, respectively. It is then possible to construct a 
general theory based on arbitrary choices for these two harmonic functions. The trivial 
choices of constant $f$ and $l$ correspond to flat spaces while less trivial choices of 
non-constant $f$ and $l$ correspond to curved spaces.

Any four-dimensional hyper-K\"ahler manifold admitting a triholomorphic isometry can be 
locally described with coordinates $q^u = x_i,t$ and a Ricci-flat metric of the 
Gibbons-Hawking form \cite{GH1,GH2,HiKLR}:
\be
ds^2 = g_{uv} dq^u dq^v = f d {\vec x\,}^2 + f^{\text{--}1} \big(dt + \vec \omega \cdot d \vec x\big)^2 \,.
\ee
This depends on a single real function $f = f(\vec x)$ of the three variables $x_i$, 
which must be harmonic and therefore satisfies the three-dimensional Laplace equation:
\be
\Delta f = 0 \,.
\ee
The three functions $\omega_i$ are determined, modulo an irrelevant 
ambiguity, by the following equation, whose integrability is guaranteed by the Laplace equation:
\be
\vec \nabla \times \vec \omega = \vec \nabla f \,.
\ee
The three closed K\"ahler forms, which satisfy $d J_i = 0$ thanks to the above equation defining 
$\omega_i$, are given by (see for instance \cite{R})
\bea
J_i = \big(d t + \vec \omega \cdot d \vec x\big) \wedge d x_i - \text{\small $\frac 12$} f \epsilon_{ijk} dx_j \wedge dx_k \,.
\eea
Finally, the isometry acts as a simple shift of the $t$ coordinate by some real parameter $\xi$,
and the associated Killing vector reads
\bea
k = \xi\, \partial_t \,.
\label{Killing}
\eea
In this parametrization, the components $g_{uv}$ and $g^{uv}$ of the metric and its inverse are 
easily worked out, and their positivity requires $f > 0$. 
The components $(J_i)_{uv}$ of the three K\"ahler forms are easily verified to satisfy the quaternionic algebra 
$(J_i)^u_{\;\;w} (J_j)^w_{\;\;v} = - \delta_{ij} \delta^u_v + \epsilon_{ijk} (J_k)^u_{\;\;v}$.
Finally, it is also straightforward to verify that this Killing vector (\ref{Killing}), whose only non-vanishing component 
is $k^t = \xi$, is triholomorphic, and that the corresponding Killing potentials $P^i$, 
defined by $\nabla_u P_i = - (J_i)_{uv} k^v$, read: 
\bea
\vec P = \xi\, \vec x \,.
\eea

Any two-dimensional special-K\"ahler manifold can be locally described with special 
complex coordinate $z$ and a metric of the following form:
\bea
ds^2 = 2\, g_{z \bar z}\, d z\, d \bar z = 2\,l\, |d z|^2 \,.
\eea
This depends on a single real function $l=l(z,\bar z)$ of the two variables $z,\bar z$, which 
must be a harmonic function corresponding to the real part of a holomorphic function related to the prepotential and 
therefore satisfies the two-dimensional Laplace equation:
\bea
\partial \bar \partial \, l = 0 \,.
\eea
In this parametrization, the unique non-trivial components of the metric and its inverse 
are given by $g_{z \bar z} = l$ and $g^{z \bar z} = l^{-1}$. Positivity of the metric requires $l > 0$.

It is worth emphasizing that the above general constructions can also be obtained in an algebraic way,
using superfields. In the hyper-K\"ahler case, one can consider an N=2 single tensor multiplet, which 
consists of a linear multiplet $L$ plus a chiral multiplet $Q$ from the N=1 perspective and automatically 
incorporates a shift symmetry \cite{LR} (see also \cite{AADT}). The most general N=2 kinetic Lagrangian 
for such a multiplet is then obtained from a potential $H=H(L,Q,\bar Q)$ which must be a harmonic function: 
$H_{LL} + H_{Q \bar Q} = 0$. After switching to a description in terms of four real scalars, one then finds 
a Gibbons-Hawking space, with $H_{LL}$ mapping to the harmonic function $f$ and ${\rm Re}\,H_{LQ}$,
${\rm Im}\,H_{LQ}$ mapping to the two non-trivial components of $\vec \omega$ (see for instance appendix C of \cite{Tz}).
In the special-K\"ahler case, one can use an N=2 vector multiplet, which consists of a chiral multiplet $\Phi$ 
plus a vector multiplet $V$ from the N=1 perspective. The most general N=2 kinetic Lagrangian for such a 
multiplet involves a potential $F=F(\Phi)$ which must be a holomorphic function: $F_{\bar \Phi} = 0$. 
Keeping complex coordinates, one then directly finds the special-K\"ahler space in the above-described form, 
with ${\rm Im}\,F_{\Phi \Phi}$ mapping to the harmonic function $l$.

Summarizing, with the above local parametrization of the two components of the scalar manifold, the 
data defining the model are the following:
\bea
\b\b g_{uv} = \left(\begin{matrix}
f + f^{\text{--}1} \omega_1^2 & f^{\text{--}1} \omega_1\, \omega_2 & f^{\text{--}1} \omega_1 \, \omega_3 & f^{\text{--}1} \omega_1 \\
f^{\text{--}1} \omega_2 \, \omega_1  & f + f^{\text{--}1} \omega_2^2 &  f^{\text{--}1} \omega_2 \, \omega_3 & f^{\text{--}1} \omega_2 \\
f^{\text{--}1} \omega_3 \, \omega_1 & f^{\text{--}1} \omega_3 \, \omega_2 & f + f^{\text{--}1} \omega_3^2 & f^{\text{--}1} \omega_ 3 \\
f^{\text{--}1} \omega_1 & f^{\text{--}1} \omega_2 & f^{\text{--}1} \omega_3 & f^{\text{--}1} \\
\end{matrix}\right) \,,\;\; 
k^u = \left(\begin{matrix}
0 \\ 0 \\ 0 \\ \xi 
\end{matrix}\right) \,, \\[1mm]
\b\b g_{z \bar z} = l \,,\;\; \rho = l \,, \\[1mm]
\b\b V = \xi^2 \Big[f^{-1} |z|^2 + \text{\small $\frac 12$}\, l^{-1} |\vec x|^2\Big] \,.
\eea

\section{Vacua and masses}

Let us now look for Poincar\'e invariant vacuum states of the above theory, 
defined by constant expectation values for the six independent scalar fields 
$q^I = x_i,t, z, \bar z$. To start, we compute the first derivative $V_I$ 
and find that $V_t = 0$ while
\bea
\b\b V_i = \xi^2 \Big[\!-\! f^{\text{--}2} |z|^2 f_i + l^{\text{--}1} x_i\Big] \,, \\
\b\b V_z = \xi^2 \Big[\!-\! \text{\small $\frac 12$}\, l^{\text{--}2} |\vec x|^2 l_z + f^{\text{--}1} \bar z\Big] \,.
\eea
For the matrix of second derivatives $V_{I \bar J}$, we instead find $V_{tt} = 0$ and $V_{it} = 0$ while
\bea
\b\b V_{ij} = \xi^2 \Big[\!-\! f^{\text{--}2} |z|^2 \big(f_{ij} - 2 f^{\text{--}1} f_i f_j\big) + l^{\text{--}1} \delta_{ij}\Big] \,, \\
\b\b V_{z \bar z} = \xi^2 \Big[\!-\! \text{\small $\frac 12$}\, l^{\text{--}2} |\vec x|^2 \big(l_{z\bar z} - 2\hspace{1pt} l^{\text{--}1} |l_z|^2\big) + f^{\text{--}1}\Big] \,, \\
\b\b V_{z z} = \xi^2 \Big[\!-\! \text{\small $\frac 12$}\, l^{\text{--}2} |\vec x|^2 \big(l_{zz} - 2\hspace{1pt} l^{\text{--}1} l_z^2\big)\Big] \,, \\[0.5mm]
\b\b V_{i z} = \xi^2 \Big[\!-\! f^{\text{--}2} \bar z\, f_i - l^{\text{--}2} x_i\, l_{z}\Big] \,.
\eea
Finally, we also need to compute the vielbein $e_I{}^P$ that allows us to locally trivialize the metric
as $g_{I \bar J} = e_I{}^P \delta_{P\bar Q}\, (e^\dagger)^{\bar Q}{}_{\bar J}$ and thus canonically normalize the scalar fields. 
One finds a block diagonal result given by:
\bea
\b\b e_i{}^p = \left(\begin{matrix}
f^{1/2} & 0 & 0 & f^{\text{--}1/2} \omega_1 \\
0 & f^{1/2} & 0 & f^{\text{--}1/2} \omega_2 \\
0 & 0 & f^{1/2} & f^{\text{--}1/2} \omega_3 \\
0 & 0 & 0 & f^{\text{--}1/2}
\end{matrix}\right) \,,\\[1mm]
\b\b e_z{}^z = l^{1/2} \,.
\eea

The possible vacua correspond to the stationary points of the potential $V$. 
The stationarity conditions $V_I = 0$ determining them are easy to analyze. We see that whenever 
$f$ or $l$ are constant and at least one of the factors of the scalar manifold is flat, stationarity implies 
vanishing values for all the fields and unbroken supersymmetry with vanishing vacuum 
energy. To get non-trivial supersymmetry-breaking stationary points, we thus need both of the functions 
$f$ and $l$ to be non-trivial and thus both factors of the scalar manifold to be curved. In that case 
the value of the fields is non-vanishing and the stationarity conditions imply that:
\bea
\b\b f_i =  f^2 l^{\text{--}1} |z|^{-2} x_i \,, \\[1mm]
\b\b l_z = 2\, l^2 \hspace{-1pt} f^{\text{--}1} |\vec x|^{-2} \bar z \,.
\eea
Using these results we can then simplify the unnormalized mass matrix $V_{I \bar J}$, and finally 
compute the physical mass matrix associated to canonically normalized fields as 
$m^2_{I \bar J} = (e^{\text{--}1})_I{}^P V_{P \bar Q}\, (e^{\text{--}1\dagger})^{\bar Q}{}_{\bar J}$. 
Although the form of $e_I{}^P$ depends on $\omega_i$ in the entries related to the would-be Goldstone
mode $t$, the final result for $m^2_{I \bar J}$ does not depend on $\omega_i$. One finds that 
$m^2_{tt} = 0$ and $m^2_{i t} = 0$ while
\bea
\b\b m^2_{ij} = \xi^2 \Big[\!-\! f^{\text{--}3} |z|^2 f_{ij} + 2\,l^{\text{--}2} |z|^{\text{--}2} x_i\, x_j + f^{\text{--}1} l^{\text{--}1} \delta_{ij}\Big] \,, \\[0.2mm]
\b\b m^2_{z \bar z} = \xi^2 \Big[\!-\! \text{\small $\frac 12$}\, l^{\text{--}3} |\vec x|^2 l_{z \bar z} + 4 f^{\text{--}2} |\vec x|^{\text{--}2} |z|^2 \! + f^{\text{--}1} l^{\text{--}1}\Big] \,,\\
\b\b m^2_{zz} = \xi^2\Big[\!-\! \text{\small $\frac 12$}\, l^{\text{--}3} |\vec x|^2 l_{zz} + 4 f^{\text{--}2} |\vec x|^{\text{--}2} \bar z^2\Big] \,, \\[0.5mm]
\b\b m^2_{i z} = \xi^2 \Big[\!-\! f^{\text{--}1/2} l^{\text{--}3/2} |z|^{\text{--}2} \bar z\, x_i  - 2\, f^{\text{--}3/2} l^{\text{--}1/2} |\vec x|^{\text{--}2} \bar z \, x_i\Big]  \,.
\eea

The unnormalized mass of the vector field $A_\mu$ can be read off from the kinetic term 
of the hypermultiplet scalars and is given by $g_{uv}k^uk^v = \xi^2 f^{\text{--}1}$. One then 
has to rescale this by $\rho^{\text{--}1} = l^{\text{--}1}$ to get the physical mass for the 
canonically normalized vector, finding
\bea
m^2_A = \xi^2 f^{\text{--}1} l^{\text{--}1} \,.
\eea

A convenient way of parametrizing the above results is to introduce an angle $\theta$ that 
controls the relative orientation of the supersymmetry breaking direction between the hyper and 
the vector sectors. To do so, we consider the ratio of the two contributions in $V$ and define
at the vacuum point:
\bea
\tan^2 \theta = \text{\small $\frac 12$} \hspace{1pt} f\,l^{\text{--}1} |z|^{\text{--}2} |\vec x|^{2} \,.
\eea
Let us also introduce the direction $v_i$ to which the vacuum point corresponds in the hypermultiplet 
field subspace, and similar the phase $\varphi$ defined by the vacuum point in the vector multiplet 
field subspace, namely:
\bea
v_i = \frac {x_i}{|\vec x|} \,,\;\; \varphi = \arg z \,.
\eea
We then parametrize the overall scale of the fields at the vacuum point by an energy scale $\Lambda$
defined as:
\be
\Lambda^2 = l \hspace{1pt} |z|^2 + \text{\small $\frac 12$} \hspace{1pt} f\hspace{1pt}  |\vec x|^2 \,.
\ee
In this way, the values of the fields are parametrized as:
\be
x_i = \sqrt{2} \hspace{1pt} f^{\text{--}1/2} \Lambda \sin \theta\, v_i \,,\;\; 
z = l^{\text{--}1/2} \Lambda \cos \theta\, e^{i \varphi} \,.
\ee
In addition, let us introduce the following dimensionless parameters associated to the second derivatives 
of the functions $f$ and $l$:
\bea
a_{ij} = f^{\text{--}1}|\vec x|^2\! f_{ij} \,,\;\; b_{z \bar z} = l^{\text{--}1} |z|^2 l_{z \bar z} \,,\;\; b_{zz} = l^{\text{--}1} z^2 l_{zz}
\eea
In this parametrization, the scalar masses $m^2_{I \bar J}$ can then be rewritten in the following 
very simple form:
\bea
\b\b m^2_{ij} = \Big[\delta_{ij} + 4\hspace{1pt} \tan^2 \theta\,v_i v_j 
- \text{\small $\frac 12$} \cot^2 \theta\, a_{ij} \Big] m^2_A \,, \label{m2ij} \\[0.2mm]
\b\b m^2_{z \bar z} = \Big[1 + 2 \cot^2 \theta - \tan^2 \theta\, b_{z \bar z}\Big] m^2_A \,,\\
\b\b m^2_{z z} = \Big[2 \cot^2 \theta - \tan^2 \theta\, b_{zz} \Big] e^{\text{--} 2 i \varphi} m^2_A \,, \label{m2zz}\\[0.2mm]
\b\b m^2_{i z} = \Big[\!-\! \sqrt{2} \big(\! \cot \theta + \tan \theta \big) v_i \Big] e^{\text{--} i \varphi}m^2_A\,. \label{m2iz}
\eea
Notice also that the vacuum energy is related to the vector mass and the scale defined by the 
expectation values of the fields:
\bea
V = \Lambda^2 m^2_A \,.
\eea

\section{Bounds on the scalar masses}

We would now like to understand what kind of values can be achieved for the scalar masses
$m^2_i$ corresponding to the eigenvalues of the mass matrix $m^2_{I \bar J}$.
To this aim, we shall take the point of view that we choose some definite point corresponding to some values 
of $\vec x$ and $z$ to be a priori the vacuum point, and then scan over all the possible forms of the functions 
$f$ and $l$ in the neighborhood of such a point. 
The condition that the chosen point should be a stationary point of $V$ fixes the values of the first derivatives 
$f_i$ and $l_z$. But the values of the functions $f$ and $l$ themselves as well as those of their second 
derivatives $f_{ij}$ and $l_{z \bar z}$, $l_{z z}$ are then arbitrary, except for the harmonicity constraints 
$\delta^{ij} f_{ij} = 0$ and $l_{z \bar z} = 0$. One may then scan over the two real parameters $f$ and $l$ 
and the seven independent real parameters among the $f_{ij}$ and $l_{z \bar z}$, $l_{zz}$, and see what 
kind of masses one can achieve. In terms of the parametrization introduced at the end of previous section, 
this means in particular that we can scan over all the possible values of $m^2_A$, which
controls the overall scale of the scalar masses, and $\theta$, $a_{ij}$, $b_{z \bar z}$, $b_{zz}$ which 
control instead the detailed form of the scalar mass matrix, with the only constraints being that:
\bea
\delta^{ij} a_{ij} = 0 \,,\;\; b_{z \bar z} = 0 \,.
\label{constraints}
\eea

To get an idea of whether it is possible or not to make all the eigenvalues positive, 
we may now look at the average values of the three blocks of the mass matrix, and 
reduce the original (4+2)-dimensional matrix to a simpler (1+1)-dimensional averaged
matrix. More precisely, taking into account that we already know that there is one null
eigenvalue in the hyper sector corresponding to the unphysical would-be Goldstone mode 
$t$ absorbed by $A_\mu$ in a Higgs mechanism, let us look at 
\bea
m^2_{\rm hh} = \text{\small $\frac 13$}\, \delta^{ij} m^2_{ij} \,,\;\;
m^2_{\rm vv} =  m^2_{z \bar z} \,,\;\;
m^2_{\rm hv} =  \sqrt{\text{\small $\frac 13$}\, \delta^{ij} m^2_{iz} m^2_{\bar \jmath \bar z}} \,.
\eea
After a straightforward computation and using the constraints (\ref{constraints}) 
imposed by the three-dimensional and two-dimensional Laplace equations 
satisfied by the functions $f$ and $l$, one finds:
\bea
\b\b m^2_{\rm hh} = \Big[1 + \text{\small $\frac 43$} \tan^2 \theta \Big] m^2_A \,, \label{m2hh} \\[1mm]
\b\b m^2_{\rm vv} = \Big[1 + 2 \cot^2 \theta \Big] m^2_A \,,  \label{m2vv} \\[-1mm]
\b\b m^2_{\rm hv} = \Big[\text{\small $\sqrt{\frac 23}$}\big(\tan \theta + \cot\theta\big) \Big] m^2_A \label{m2hv} \,.
\eea
We see that as a result of the constraints imposed by $N=2$ supersymmetry, and in particular 
(\ref{constraints}), these average blocks are almost completely fixed, the only leftover parameter 
being the angle $\theta$ controlling the relative strength of the hyper- and vector multiplet sectors 
in the supersymmetry breaking process. 

The first, qualitative information that we can extract from the knowledge of the above averaged 
blocks concerns the sign of the eigenvalues $m^2_{i}$. Some simple linear algebra shows 
that the full six-dimensional mass matrix $m^2_{IJ}$ can be positive definite only if the 
two-dimensional averaged mass matrix is also positive definite. This is the case 
if $m^2_{\rm hh} > 0$, $m^2_{\rm vv} > 0$ and $m^2_{\rm hh} m^2_{\rm vv} - m^4_{\rm hv} > 0$.
It is straightforward to check that all these three conditions are always satisfied by the expressions 
(\ref{m2hh}), (\ref{m2vv}) and (\ref{m2hv}), and this for any possible value of the angle $\theta$.
This suggests that it is a priori possible to adjust the parameters $a_{ij}$ and $b_{z \bar z}$, $b_{zz}$
subject to the constraints (\ref{constraints}) in such a way to make all the eigenvalues $m^2_i$ positive.

The second, quantitative information that we can extract from the knowledge of the above averaged 
blocks concerns the size of the eigenvalues $m^2_{i}$. Since for a given $\theta$ all the averaged 
blocks are bounded, relative to the overall scale $m^2_A$, it is clear that the eigenvalues $m^2_i$ 
must also be bounded to lie in a certain interval, again relative to the overall scale $m^2_A$. 
More precisely, there must be an upper bound $m^2_-$ 
on how large the smallest $m^2_i$ can be, and also a lower bound $m^2_+$ on how small the largest 
$m^2_i$ can be. Through some simple linear algebra, one can show that these bounds $m^2_\pm$ are 
in fact simply the two eigenvalues of the two-dimensional matrix formed by the averaged mass blocks 
$m^2_{\rm hh}$, $m^2_{\rm vv}$ and $m^2_{\rm hv}$, and are thus given by:
\bea
m^2_{\pm} = \text{\small $\frac 12$} \big(m^2_{\rm hh} + m^2_{\rm vv}\big)
\pm \sqrt{\text{\small $\frac 14$} \big(m^2_{\rm hh} - m^2_{\rm vv} \big)^2 + m^4_{\rm hv}} \,.
\eea
Using the fact that $m^2_{\rm hh}> 0$,  $m^2_{\rm vv} > 0$ and $m^2_{\rm hh} m^2_{\rm vv} - m^4_{\rm hv} > 0$,
one can then infer the following bounds, which can be derived by studying the necessary conditions for the matrix 
$m^2_{IJ} - m^2_\pm \delta_{IJ}$ to be negative or positive definite obtained after averaging and reducing to a 
two-dimensional matrix:
\bea
\a\a \min \big\{m^2_i\big\} \le m^2_{\rm -} \le \min \big\{m^2_{\rm hh}, m^2_{\rm vv} \big\} \,,\\
\a\a \max \big\{m^2_i\big\} \ge m^2_{\rm +} \ge \max \big\{m^2_{\rm hh}, m^2_{\rm vv} \big\} \,.
\eea
A simple computation shows that the quantities $m^2_\pm$ are given by
\bea
\b\b m^2_{\pm} = \bigg[1 + \cot^2 \theta + \text{\small $\frac 23$} \tan^2 \theta  \nn \\
\b\b \hspace{40pt} \pm\, \sqrt{\text{\small $\frac 23$} \cot^2 \theta + \cot^4 \theta 
+ \text{\small $\frac 23$} \tan^2 \theta + \text{\small $\frac 49$} \tan^4 \theta}\, \bigg] m^2_A \,.
\eea
One can easily verify that $m^2_+ > 0$ and $m^2_- > 0$ for any value of $\theta$, as already implied 
by the analysis of the previous paragraph. One can, however, also study more quantitatively what happens 
when $\theta$ is varied. $m^2_-$ starts from a local minimum for $\theta = 0$ with 
value $\frac 23\, m^2_A$, then goes through an absolute maximum for $\theta= \frac {\pi}4$ with value 
$m^2_A$ and finally reaches a local minimum for $\theta = \frac {\pi}2$ with value $\frac 12\, m^2_A$. 
$m^2_+$ starts from a maximum for $\theta = 0$ with infinite value, goes through a minimum at 
$\theta \simeq 0.83$ (close to $\theta = \frac {\pi}4$) with value $4.27\,m^2_A$ (close to $\frac {13}3\,m^2_A$) 
and then reaches again a maximum at $\theta = \frac {\pi}2$ with infinite value. We then conclude that:
\bea
\min \big\{m^2_{i}\big\} \le m^2_A \,,\;\; 
\max \big\{m^2_{i}\big\} \gsim \text{\small $4.27$}\, m^2_A\,.
\eea
This result suggests that it should not only be possible to make all the mass eigenvalues $m^2_{i}$ positive, 
but actually all greater than or equal to $m^2_A$. In other words, it should be possible to achieve a genuinely 
metastable supersymmetry breaking vacuum with sizable masses for scalar fluctuations by adjusting the 
parameters of the model. 

Note that the cases of theories with just one hypermultiplet or just one vector multiplet 
can formally be obtained as special cases of the more general situation studied here, by 
taking the limits $l \to + \infty$, $z \to z_0$ and $f \to + \infty$, $\vec x \to \vec x_0$, respectively.  
In those two limits one thus gets $\theta \to 0$ and $\theta \to \frac {\pi}2$, respectively, but also $m_A \to 0$ 
and $\Lambda \to +\infty$ with $V \to \text{finite}$, in both cases. One then correctly recovers the 
vanishing upper bound for the smallest mass that leads to a no-go theorem in those cases 
\cite{GRLS,MANY}, as a consequence of the vanishing of the trace of the relevant mass matrix block.

\section{Existence of metastable vacua}

The final question that we need to address is whether the full five-dimensional non-trivial
part of the mass matrix $m^2_{I \bar J}$ defined by eqs.~(\ref{m2ij})-(\ref{m2iz}) can really be made
positive definite by a suitable choice of the parameters $a_{ij}$ and $b_{z \bar z},b_{zz}$, subject to 
the constraints (\ref{constraints}). We saw that the necessary conditions for 
this to be possible that come from the study of the two-dimensional matrix obtained by averaging 
over each of the hyper and vector subsectors are satisfied for any value of $\theta$, so the question 
is more precisely whether for any given $\theta$ and $v_i,\varphi$ it is possible or not to make all 
the $m^2_i$ positive through a suitable choice of the parameters $a_{ij}$ and $b_{z \bar z},b_{zz}$. 
The answer to this question is yes, and in fact it turns out that one can always saturate the 
bounds defined by $m^2_+$ or $m^2_-$ by suitably adjusting $a_{ij}$ and $b_{z \bar z},b_{zz}$. 
An intuitive argument for this is as follows. 
Due to the restriction that $\delta^{ij} a_{ij} = 0$ and $b_{z \bar z} = 0$, the average of the 
eigenvalues of the two diagonal blocks of the mass matrix are fixed and cannot be changed. 
Moreover, the off diagonal block is also fixed and independent of the above parameters.
As a result, there is certain amount of level-repulsion between the two groups of eigenvalues 
that the diagonal blocks would have on their own, and the average value of all the eigenvalues 
of the full matrix is also fixed. It is then clear that changing $a_{ij}$ and $b_{\alpha \beta}$ can 
only affect the spread of the eigenvalues around what is dictated by the two-dimensional matrix 
obtained by averaging over the directions defining each subsector, and as a consequence 
it is possible to choose $a_{ij}$ and $b_{z \bar z},b_{zz}$ in such a way as to saturate the bounds 
defined by $m^2_+$ or $m^2_-$. 

Let us illustrate the above statement with an explicit example of metastable supersymmetry breaking 
vacuum. For simplicity, we choose the vacuum point to be defined by values of the fields in the 
maximally symmetric direction such that 
\bea
\theta = \text{\small $\frac {\pi}4$} \,,\;\; v_i = \text{\small $\sqrt{\frac 13}$} \,,\;\; \varphi = 0 \,.
\eea
The values of the functions $f$ and $l$ at such a point are arbitrary and are mapped to 
arbitrary values for the scales $m_A$ and $\Lambda$. The first derivatives of the functions 
$f$ and $l$ at such a point are instead completely fixed by the requirement that the stationarity 
conditions should be satisfied. Finally, the second derivatives of the functions $f$ and $l$ 
at such a point are arbitrary and are mapped to arbitrary values for the dimensionless parameters 
$a_{ij}$ and $b_{z \bar z},b_{zz}$. For generic values of the latter, we then get the following structure 
for the three blocks of the mass matrix:
\bea
\b\b m^2_{ij} = \left(\begin{matrix}
\displaystyle{\text{\small $\frac 73$}} - \displaystyle{\text{\small $\frac 12$}} \,a_{11} & 
\displaystyle{\text{\small $\frac 43$}} - \displaystyle{\text{\small $\frac 12$}}\,a_{12} & 
\displaystyle{\text{\small $\frac 43$}} - \displaystyle{\text{\small $\frac 12$}}\,a_{13} \\[3mm]
\displaystyle{\text{\small $\frac 43$}} - \displaystyle{\text{\small $\frac 12$}}\,a_{12} & 
\displaystyle{\text{\small $\frac 73$}} - \displaystyle{\text{\small $\frac 12$}}\,a_{22} & 
\displaystyle{\text{\small $\frac 43$}} - \displaystyle{\text{\small $\frac 12$}}\,a_{23} \\[3mm]
\displaystyle{\text{\small $\frac 43$}} - \displaystyle{\text{\small $\frac 12$}}\,a_{13} & 
\displaystyle{\text{\small $\frac 43$}} - \displaystyle{\text{\small $\frac 12$}}\,a_{23} & 
\displaystyle{\text{\small $\frac 73$}} - \displaystyle{\text{\small $\frac 12$}}\,a_{33} \\
\end{matrix}\right) m^2_A \,,
\eea
\vspace{-8pt}
\bea
\b\b m^2_{\alpha \bar \beta} = \left(\begin{matrix}
3 - b_{z \bar z}& 2 - b_{z z}\\[2mm]
2 - b_{\bar z \bar z} & 3 - b_{z \bar z}\\
\end{matrix}\right) m^2_A \,,
\eea
\vspace{-8pt}
\bea
\b\b m^2_{i\bar \beta} = \left(\begin{matrix}
- \displaystyle{\text{\small $\sqrt{\frac 83}$}} & - \displaystyle{\text{\small $\sqrt{\frac 83}$}}\\[3mm]
- \displaystyle{\text{\small $\sqrt{\frac 83}$}} & - \displaystyle{\text{\small $\sqrt{\frac 83}$}} \\[3mm]
- \displaystyle{\text{\small $\sqrt{\frac 83}$}} & - \displaystyle{\text{\small $\sqrt{\frac 83}$}} \\
\end{matrix}\right) m^2_A \,.
\eea
Recalling the constraints $\delta^{ij} a_{ij} = 0$ and $b_{z \bar z} = 0$, in this case 
we have
\bea
m^2_{\rm hh} = \text{\small $\frac 73$}\,m^2_A \,,\;\; m^2_{\rm vv} = 3\, m^2_A \,,\;\;
m^2_{\rm hv} = \text{\small $\sqrt{\frac 83}$}\, m^2_A \,.
\eea
and
\bea
m^2_- = m^2_A \,,\;\; m^2_+ = \text{\small $\frac {13}3$}\,m^2_A\,.
\eea
We can finally make some definite choice for the parameters $a_{ij}$ and $b_{z \bar z}, b_{zz}$
and compute the mass eigenvalues $m^2_i$ explicitly. As expected, by choosing 
appropriate values for these parameters it is possible to make all the $m^2_i$ positive, 
but at least one of these is always lighter that $m^2_- = m^2_A$ and one 
is always heavier than $m^2_+ = \frac {13}3\, m^2_A$. A very simple working example 
of the above type is obtained by making the following choice of parameters:
\bea
\b\b a_{ij} = 0\,,\;\; b_{z \bar z}, b_{zz}= 0 \,.
\eea
In this case, the five non-trivial eigenvalues of the full mass matrix can be computed 
analytically and are found to be:
\bea
m^2_{i} \a=\a \big\{1,1,1,1,9 \big\}\, m^2_A\,.
\eea
These are all positive, and the vacuum is thus metastable. We moreover see that in 
this simple example the upper bound on the lightest mass is saturated. 

\section{Generalization to supergravity}

The results that we have derived here in the context of rigid supersymmetry can 
be generalized to local supersymmetry. To do so, one needs to consider a generic 
supergravity theory with one hyper- and one vector multiplet. The hypermultiplet 
sector is described by a four-dimensional quaternionic-K\"ahler manifold with 
negative Ricci curvature set by the Planck scale, and this must again 
admit a triholomorphic isometry. Fortunately, the most general space with these 
properties is also known and goes under the name of the Przanowski-Tod space \cite{P,T}. 
This is the Ricci-curved generalization of the Gibbons-Hawking space, 
and is based on a function of three variables satisfying the non-linear three-dimensional 
Toda equation, rather than the linear three-dimensional Laplace equation. The vector multiplet sector is 
instead described by a two-dimensional local-special-K\"ahler manifold. This can also 
be described in a completely general way. The new feature is again a deformation 
in the structure of the curvature by effects linked to the Planck scale.

A detailed analysis of the structure of the mass matrix, the bounds that can be put 
on its eigenvalues and the constraints on the possibility of achieving metastable 
de Sitter vacua in this kind of theories can be performed by using the technology 
described in \cite{CSS1} and will be presented elsewhere \cite{CSS2}, along with 
some explicit examples. It is however clear that the existence of metastable 
supersymmetry-breaking vacua in the rigid limit directly implies also the existence of metastable 
supersymmetry breaking de Sitter vacua in supergravity. This shows that Fayet-Iliopoulos terms 
and non-Abelian gauge symmetries are not necessary ingredients to achieve metastable 
supersymmetry breaking even within supergravity, again contrary to the common lore in the literature
and in particular the claim of \cite{FTV}.  The only subtle point concerns the values 
of the cosmological constant $V$ and the gravitino mass $m_{3/2}$ that can be 
compatible with metastability. It is obvious that in the limit where $V \gg m^2_{3/2} M^2_{\rm Pl}$, 
gravitational effects on supersymmetry breaking and on the masses are small and it must 
therefore be possible to achieve metastable de Sitter vacua exactly as in the rigid case. On the other
hand, in the limit where $V \ll m^2_{3/2} M^2_{\rm Pl}$, gravitational effects on supersymmetry 
breaking and on the masses are sizable and the possibility of achieving metastable de 
Sitter vacua must be carefully reinvestigated. The quantitative question that one then 
has to deal with consists in understanding for which range of values of the dimensionless 
ratio $V/(m^2_{3/2} M_{Pl}^2)$ metastability can be achieved. This is a particularly relevant 
question, since small and large values of the above parameter are needed in applications to 
particle physics and inflation, respectively.

\section{Conclusions}

In this letter, we have demonstrated that metastable spontaneous breaking of global 
N=2 supersymmetry is possible even in very simple theories that do not involve 
Fayet-Iliopoulos terms or non-Abelian gaugings. We then 
argued that the same qualitative result also holds true in the presence of 
gravity, although the relative size of the cosmological constant and the gravitino mass
allowing for metastable vacua might be constrained and remains to be analyzed. 

To conclude, let us compare our findings with the general statement in 
\cite{AB} that N=2 theories admitting an $SU(2)_R$ symmetry and a supercurrent 
conservation law based on a linear superconformal anomaly multiplet cannot 
spontaneously break supersymmetry. Our examples of N=2 theories possessing
metastable supersymmetry-breaking vacua have a priori no $SU(2)_R$ symmetry, 
since the generic Gibbons-Hawking manifolds we considered do not admit an 
isometry group that could contain this. We believe that this is the reason why they 
evade the result of \cite{AB}. As a consistency check of this interpretation, we 
verified that in the special models built on spaces with a larger isometry group, 
such as flat space and the Eguchi-Hanson manifold, there are in fact no 
supersymmetry-breaking vacua. Nevertheless, it would be interesting to 
understand whether or not our models admit a linear superconformal anomaly 
multiplet coping with the potential problems emphasized in \cite{KS,DKS}. 

\vskip 20pt
\noindent
{\Large \bf Acknowledgements}
\vskip 10pt
\noindent
We are grateful to M.~Gomez-Reino and  J.~Louis for earlier collaboration 
on various related issues, and to I.~Antoniadis and M.~Buican for extensive 
discussions about their work \cite{AB}. We also thank L.~Alvarez-Gaum\'e, 
F.~Catino, J.-P.~Derendinger, S.~Ferrara, J.~Fine, T.~Hausel and J.-C.~Jacot 
for useful discussions. The research of C.~S and P.~S. is supported by the 
Swiss National Science Foundation under the grant PP00P2-135164.

\end{document}